\documentclass{ws-procs961x669}            

\def\Lie{\pounds}

\begin{document}

\title{Gravitational Energy and the Gauge Theory Perspective}

\author{Chiang-Mei Chen$^{1,2,\dag}$ and James M. Nester$^{1,2,3,4,*}$}

\address{$^1$Department of Physics, National Central University, Chungli 32001, Taiwan\\
$^2$Center for Mathematical and Theoretical Physics, National Central University\\ Chungli 32001, Taiwan\\
$^3$Graduate Institute of Astronomy, National Central University,\\ Chungli 32001, Taiwan\\
$^4$Leung Center for Cosmology and Particle Astrophysics,\\
National Taiwan University, Taipei 10617, Taiwan\\
$^*$E-mail: nester@phy.ncu.edu.tw\\
$^\dag$E-mail: cmchen@phy.ncu.edu.tw}

\begin{abstract}
Gravity, and the puzzle regarding its energy, can be understood from a gauge theory perspective.
Gravity, i.e., dynamical spacetime geometry, can be considered as a local gauge theory of the symmetry group of Minkowski spacetime: the Poincar\'e group.  The dynamical potentials of the Poincar\'e gauge theory of gravity  are the frame and the metric-compatible connection.  The spacetime geometry has in general both curvature and torsion.  Einstein's general relativity theory is a special case.
Both local gauge freedom and energy are clarified via the Hamiltonian formulation.
We have developed a covariant Hamiltonian formulation. The Hamiltonian boundary term gives covariant expressions for the quasi-local energy, momentum and angular momentum.  A key feature is the necessity to choose on the boundary a non-dynamic reference.  With a best matched reference one gets good quasi-local energy-momentum and angular momentum values.
\end{abstract}


\bodymatter

\begin{quote}\textit{Dedication}: To the memory of Prof.~Yi-Shi Duan, who inspired many students and did pioneering work on many subjects---including two that have been our interest: the gauge theory formulation of gravity and identifying good expressions for the energy-momentum and angular momentum of gravitating systems.\cite{Duan:1986if,Duan:1988grg,Duan:1996cpl,Duan:1998uu,Feng:1995iv,Feng:1996cpl,Feng:1999ij,
Liu:2007iy}
 Prof.~Duan  did much to encourage  attention to these important topics.
\end{quote}

\section{Introduction}
The evolution of a generally covariant theory is under-determined.  This first became an issue  in connection with Einstein's gravity theory, general relativity (GR).  One consequence is that gravitational energy has no proper localization. Trying to clarify this fact led ultimately to the gauge theories of physical interactions.

GR with general covariance is the premier gauge theory. The consequences of this,
especially regarding  gravitational energy and under-determined evolution, were long perplexing.    The Hamiltonian approach clarifies these issues.  Gravity can be understood as a gauge theory of the local Poincar\'e symmetries of spacetime.

As noted above, Prof.~Duan was much concerned with these issues.  Here we present an introduction to our work in this area.  It will be noticed that we use many of the same ideas that were used by Prof.~Duan.  The distinctive features of our approach is
that we use the first order Lagrangian and the Hamiltonian formulations, moreover, we always favor a representation in terms of differential forms.

\section{Some Historical Background}

Dynamical equations obtained from a variational principle formerly had deterministic Cauchy initial value problems, but in GR there appears a differential identity connecting the evolution equations, they were not independent and could not give uniquely determined evolution---this is \emph{the essence of gauge theory}.
This type of indeterminism was later found to be best addressed via the Hamiltonian approach.\cite{Dirac64}

\subsection{Automatic Conservation of the Source and Gauge Fields}

 In 1916 Einstein showed that local coordinate invariance plus his field equations gives material energy momentum conservation, without using the matter field equations (see Doc.~41 in Vol.~6 of Ref.~\refcite{CPAE}).
This is referred to as \emph{automatic conservation of the source\/} (see section 17.1 in Ref.~\refcite{MTW});  it uses  a Noether second theorem local (gauge) symmetry type of argument to obtain current conservation.
Hermann Weyl argued in this way for the electromagnetic current in his  papers of 1918 (the name  \emph{gauge theory\/} comes from this work) and 1929,\footnote{An English translation of Weyl's seminal papers can be found in Ref.~\refcite{Dawn}.} whereas modern field theory generally uses Noether's first theorem for current conservation.

The \emph{essence\/} of gauge theory is \emph{a local symmetry\/}, consequently:
(i) a differential identity,
(ii)  under-determined evolution,
(iii) restricted type of source coupling,
(iv) automatic conservation of the source.
Yang-Mills is only one special type.
Our gauge approach to gravity does not try to force it into the Yang-Mills mold, but rather simply recognizes the natural local symmetries associated with the spacetime geometry.

\section{Noether's 1918 Contribution}

One word well describes 20th century physics: \emph{symmetry}.
Most of the theoretical physics ideas involved symmetry---essentially they are applications of Noether's two theorems.\cite{KS-Noether}
The first associates conserved quantities with global symmetries.
The second concerns \emph{local symmetries}: it is \emph{the foundation of the modern gauge theories}.

Why did Noether make her investigation?
She was a mathematician; her interest was not physics. At the time she was assisting Hilbert and Klein, especially in connection with the puzzling issue of energy in GR.
After presenting her two famous theorems she uses them to draw the conclusion that clarifies the situation.\cite{KS-Noether}

Her result regarding \emph{``the lack of a proper law of energy''} applies not just to Einstein's GR, but \textit{to all geometric theories of gravity}. For gravitating systems there is no well-defined \textit{local} energy-momentum density.
The modern view is that energy-momentum is not ``local'' (i.e., meaningful as a density at a point) but rather \emph{quasi-local}---associated with a closed 2-surface.\cite{Sza09}

\section{Energy-momentum Pseudotensors and the Hamiltonian}

The {Einstein Lagrangian} differs from Hilbert's by a total divergence:
\begin{eqnarray}
2\kappa{\cal L}_{\rm E}(g_{\alpha\beta}, \partial_\mu g_{\alpha\beta}):=
-\sqrt{-g}g^{\beta\sigma} \Gamma^\alpha{}_{\gamma\mu} \Gamma^\gamma{}_{\beta\nu}\delta^{\mu\nu}_{\alpha\sigma}
\equiv
\sqrt{-g}R-\hbox{div}.
\end{eqnarray}
From this Einstein constructed the associated canonical energy-momentum density, now known as the \emph{Einstein pseudotensor\/}:\footnote{It is not a \emph{proper} tensor.}
\begin{equation}\label{EpseudoT}
\mathfrak{t}_{\rm E}^\mu{}_\nu := \delta^\mu_\nu{\cal L}_{\rm E} - \frac{\partial{\cal L}_{\rm E}}{\partial \partial_\mu g_{\alpha\beta}} \partial_\nu g_{\alpha\beta}.
\end{equation}
Using the Einstein equation $\sqrt{-g}G^\mu{}_\nu=\kappa\mathfrak{T}^\mu{}_\nu$ one gets a
 conserved total energy-momentum:
\begin{equation}
\partial_\mu (\mathfrak{T}^\mu{}_\nu+\mathfrak{t}_{\rm E}^\mu{}_\nu)=0,
\qquad\Longleftrightarrow\qquad
\sqrt{-g}G^\mu{}_\nu+\kappa\mathfrak{t}_{\rm E}^\mu{}_\nu=\partial_\lambda \mathfrak{U}^{[\mu\lambda]}{}_\nu.
\end{equation}
A good form for the superpotential $\mathfrak{U}$ was found only much later by Freud in 1939:\cite{Freud}
$\mathfrak{U}_{\rm F}^{\mu\lambda}{}_\nu:=-\mathfrak{g}^{\beta\sigma}
\Gamma^\alpha{}_{\beta\gamma}\delta^{\mu\lambda\gamma}_{\alpha\sigma\nu}
$.
Other pseudotensors
 similarly follow from different superpotentials.  They are all inherently coordinate reference frame dependent.
Thus there are two big problems: (1) \emph{which pseudotensor?} (2) \emph{which reference frame?}
The Hamiltonian approach, as we shall explain, has answers.

With constant $Z^\mu$, the energy-momentum within a region is
\begin{eqnarray}
-Z^\mu P_\mu(V) &:=& - \int_V Z^\mu ({{\mathfrak{T}}^\nu{}_\mu+{\mathfrak{t}}^\nu{}_\mu}) \sqrt{-g}d^3\Sigma_\nu
\nonumber\\
&\equiv& \int_V \left[ Z^\mu \sqrt{-g} \left( \frac1{\kappa} G^\nu{}_\mu - T^\nu{}_\mu \right) - \frac1{2\kappa} \partial_\lambda \left( Z^\mu {\mathfrak{U}^{\nu\lambda}{}_\mu} \right) \right] d^3\Sigma_\nu
\nonumber\\
&\equiv& \int_V Z^\mu {\cal H}^{\rm GR}_\mu + \oint_{S=\partial V} {\cal B}^{\rm GR}(Z) \equiv H(Z, V). \label{basicHam}
\end{eqnarray}
${\cal H}^{\rm GR}_\mu$ is the well known covariant expression for the \textit{Hamiltonian density}.
  The {\em boundary term\/} 2-surface integral
 is determined by the superpotential.
The value of the
pseudotensor/Hamiltonian is thus \emph{quasi-local}, from just the boundary term, since by the initial value constraints the spatial volume integral vanishes.

\section{The Hamiltonian Approach}

Noether's work
  can be combined with the Hamiltonian formulation.
In Hamiltonian field theory, \emph{the conserved currents are the generators of the associated symmetry}.
For local spacetime ``translations''  (i.e., infinitesimal diffeomorphisms), the associated  current expression (i.e., the energy-momentum density) \emph{is\/} the Hamiltonian density---the canonical generator of spacetime displacements.
   Because it can be varied it gives a handle on the conserved current ambiguity.   As we will explain, the Hamiltonian variation gives information that tames the ambiguity in the boundary term---namely boundary conditions. In this way problem 1 is under control.  Pseudotensor values are values of the Hamiltonian with certain boundary conditions.\cite{CNC99}

The Hamiltonian approach reveals certain aspects of a theory.
The constrained Hamiltonian formalism was developed by Dirac\cite{Dirac64} and by Bergmann and coworkers.  It was applied to GR by Pirani, Schild and Skinner\cite{PSS} and by Dirac\cite{Dirac58}.  Later the ADM approach\cite{ADM} became dominant.
For the Poincar\'e gauge theory of gravity (PG) the Hamiltonian approach was developed by Blagojevi\'c and coworkers.\cite{BN83}

\section{Gauge and Geometry}

For a good account of the early history of gauge theory see Ref.~\refcite{Dawn}.  Einstein's theory of general relativity (GR) with its principle of \textit{general covariance} was the first recognized gauge theory, the first physical theory where a local gauge symmetry was understood from the beginning as playing a major role.  Inspired by GR, Weyl,\cite{Weyl1918,Weyl1929} in his seminal works that developed a gauge theory of electrodynamics, identified the key features of all gauge theories.

In 1916 Einstein showed that local coordinate invariance plus his field equations gives material energy momentum conservation, without using the matter field equations (see Doc.~41 in Vol.~6 of Ref.~\refcite{CPAE}).
This is referred to as \emph{automatic conservation of the source\/} (see section 17.1 in Ref.~\refcite{MTW});  it uses  a Noether second theorem local (gauge) symmetry type of argument to obtain current conservation.
Hermann Weyl argued in this way for the electromagnetic current in his  papers of 1918 (the name  \emph{gauge theory\/} comes from this work) and 1929,\footnote{For an English translation of Weyl's  papers see Ref.~\refcite{Dawn}.} whereas modern field theory generally uses Noether's first theorem for current conservation.

The \emph{essence\/} of gauge theory is \emph{a local symmetry\/}, consequently:
(i) a differential identity,
(ii)  under-determined evolution,
(iii) restricted type of source coupling,
(iv) automatic conservation of the source.

Yang-Mills is only one special type of gauge theory.
Our gauge approach to gravity does not try to force it into the Yang-Mills mold, but rather simply recognizes the natural symmetries of spacetime geometry.

Gravity viewed explicitly as a gauge theory was pioneered by Utiyama (1956, 1959), Sciama (1961) and Kibble (1961).
For accounts of gravity as a spacetime symmetry gauge theory, see Hayashi \& Shirifuji\cite{HayashiShirafuji}, Hehl and coworkers\cite{HHKN,Hehl80,MAG,GFHF96}, Mielke\cite{MieE87} and Blagojevi\'c\cite{Blag02}.  A comprehensive reader with  summaries, discussions and reprints has recently appeared.\cite{BlagHehl}
GR can be seen as the original gauge theory:  the first physical theory where a local gauge freedom (general covariance) played a key role.
Although the electrodynamics potentials with their gauge freedom were known long before GR
 yet this gauge invariance was not seen as having any important role in connection with the nature of the interaction, the conservation of current,  or a differential identity---until the seminal work of Weyl, which post-dated (and was inspired by) GR.

 We also should draw attention to the parallel developments of the concept of a  connection in geometry by
   Levi-Civita, Weyl, Schouten, Cartan, Eddington, and others.
{Riemann-Cartan geometry} (with a metric and a metric compatible connection, having both curvature and torsion) is the most appropriate for a dynamic spacetime geometry theory: its local symmetries are just those of the local Poincar\'e group.
The conserved quantities, energy-momentum and angular momentum/center-of-mass momentum are associated with  the  Minkowski spacetime symmetry, i.e., the Poincar\'e group.

\section{Geometry: kinematics and dynamics}

For general dynamical geometry (\emph{metric-affine gravity}, MAG),\cite{MAG} the geometric potentials can be taken as the \textit{metric} $g_{\mu\nu}$, the \textit{co-frame} one-form $\vartheta^\mu$ and the (\emph{a priori} independent) \textit{connection} one-form $\Gamma^\alpha{}_\beta$.  The respective field strengths are
\begin{align}
 Q_{\mu\nu}&:=&\!\!\!\!-Dg_{\mu\nu}&:=&\!-dg_{\mu\nu}+\Gamma^\gamma{}_\mu g_{\gamma\nu}
+\Gamma^\gamma{}_\nu g_{\mu\gamma},\quad &\hbox{non-metricity one-form}& \\
 T^\alpha&:=&\!\!\!\!D\vartheta^\alpha&:=&\!d\vartheta^\alpha+\Gamma^\alpha{}_\beta\wedge\vartheta^\beta,\quad &\hbox{torsion two-form}& \\
 R^\alpha{}_\beta&:=&\!\!\!\!D\Gamma^\alpha{}_\beta&:=&\!d\Gamma^\alpha{}_\beta+\Gamma^\alpha{}_\gamma\wedge\Gamma^\gamma{}_\beta,
 \quad&\hbox{curvature two-form}&
\end{align}
which have the respective \textit{Bianchi identities}:
\begin{align}
&&DQ_{\mu\nu}&\equiv&-D^2g_{\mu\nu}&\equiv& R_{\mu\nu}+R_{\nu\mu},&& && &&\\
&&DT^\alpha&\equiv&D^2\vartheta^\alpha&\equiv& R^\alpha{}_\beta\wedge\vartheta^\beta,&& && &&\\
&&DR^\alpha{}_\beta&\equiv& D^2\Gamma^\alpha{}_\beta&\equiv&0.&& && &&\label{Bianchi2Id}
\end{align}

Second order field equations for dynamical geometry can be obtained by varying the potentials independently in a Lagrangian 4-form:\footnote{For a manifestly covariant formulation without any frames or components see Ref.~\refcite{Nester08}.}
\begin{equation}
{\cal L}={\cal L}(g_{\mu\nu},\vartheta^\mu,\Gamma^\alpha{}_\beta; Q_{\mu\nu}, T^\mu, R^\alpha{}_\beta).
\end{equation}
Because of the diffeomorphism and local frame gauge symmetries, the resultant field equations will satisfy the associated differential identities.

An alternative is a \textit{first-order} Lagrangian of the form
\begin{equation}
{\cal L}^1=Dg_{\mu\nu}\wedge \pi^{\mu\nu}+D\vartheta^\mu\wedge \tau_\mu+D\Gamma^\alpha{}_\beta\wedge\rho_{\alpha}{}^\beta+\Lambda(g,\vartheta,\Gamma;\pi,\tau,\rho),
\end{equation}
for which independent variations of the potentials and their associated conjugate momentum fields leads to pairs of \textit{first-order} equations.
Such a formulation has some advantages.  The connection constraints of \textit{metric compatible}, \textit{symmetic} or \textit{teleparallel} can be easily imposed simply by choosing the potential $\Lambda$ to be independent of the associated conjugate momentum field; then the field equation obtained from variation with respect to the conjugate momentum leads, respectively, to vanishing non-metricity, torsion or curvature.

Another advantage is that a first-order formulation allows for the construction of a covariant Hamiltonian formulation, as we shall explain below.

 The frame can always be restricted to be \textit{orthonormal}, then the metric coefficients are \textit{constants} and one can then eliminate the metric as a dynamical variable; the frame will still have \textit{local Lorentz} gauge freedom. The infinitesimal local symmetry group is then local infinitesimal Lorentz frame gauge transformations plus infinitesimal diffeomorphisms, i.e., local displacements of the spacetime point.  This can be recognized as the local Poincar\'e group, which can be viewed as acting on the local observer and his frame.  Thus dynamical geometry gravitational theories are \textit{naturally} local gauge theories of the symmetry group of Minkowski spacetime: \textit{Poincar\'e gauge theory}.

It is customary to use the term \textit{Poincar\'e gauge theory of gravity} (PG) to be restricted to the case where the connection is \textit{metric compatible}.  The geometry is then Riemann-Cartan, with curvature and torsion.  Although one could still use the more general metric-affine formulation with the metric compatible constraint imposed by using the metric's conjugate momentum as a  Lagrange multiplier, it is more revealing---and more efficient---to restrict to using an orthonormal frame and drop the metric and its momentum as dynamical variables. In an orthonormal frame with a metric compatible connection both $\Gamma^{\alpha\beta}$ and $R^{\alpha\beta}$ are anti-symmetric (Lorentz Lie algebra valued forms).

The MAG and PG equations and Noether differential identities along with the interaction with source fields are rather lengthy; they have been presented and discussed in detail elsewhere.\cite{MAG,Nester:2016rsy,GR100}

\section{The Poincar\'e Gauge Theory of Gravity}

The standard PG Lagrangian density has a quadratic field strength form:\footnote{$\kappa:=8\pi G/c^4$ and $\varrho^{-1}$ has the dimensions of action.
 $\Lambda$ is the cosmological constant.}
\begin{eqnarray}\label{quadraticL}
{\cal L}_{\rm PG} \sim \frac{1}{\kappa}\left(\Lambda+ \text{curvature}
  +\text{torsion}^2\right) + \frac{1}{\varrho}\,\text{curvature}^2\,,
\end{eqnarray}
Varying $\vartheta,\Gamma$ gives quasi-linear 2nd order dynamical equations for the potentials: 
\begin{eqnarray}
\kappa^{-1}(\Lambda+ \hbox{curv} + D \hbox{ tor} + \hbox{tor}^2)+ \varrho^{-1}\hbox{curv}^2&=& \hbox{energy-momentum},\qquad\\
\kappa^{-1}\hbox{tor}+\varrho^{-1} D\hbox{ curv}&=& \hbox{spin}.
\end{eqnarray}

In complete detail the general quadratic PG Lagrangian 4-form is~\cite{BHN}
\begin{eqnarray}\nonumber
  {\cal L}_{\rm PG}  &=&
  \frac{1}{2\kappa}\Bigl(a_0 R\eta+b_0 X\eta
    -2\Lambda\eta +
     \textstyle\sum
    \limits_{I=1}^{3}a_{I}{}^{(I)}T^\alpha\wedge *{}^{(I)}
    T_\alpha\Bigr)\nonumber\\
  &&  + \frac{1}{{\kappa}}\left( {\sigma}_{1}{}
^{(1)}T^{\alpha}\wedge\, ^{(1)}T_{\alpha} + {\sigma}_{2}{}
^{(2)}T^{\alpha}\wedge{} ^{(3)}T_{\alpha}\right)\cr
     & & -\frac{1}{2\varrho} 
  \Bigl(\textstyle\sum\limits_{I=1}^{6}w_{I}{}^{(I)}R^{\alpha\beta}\wedge*{}^{(I)}
  R_{\alpha\beta}\Bigr)\,\nonumber \\ 
&&-\frac{1}{2{\varrho}}\left(\ {\mu}_{1}{} ^{(1)}R^{\alpha\beta}\wedge{}
^{(1)}R_{\alpha\beta} + {\mu}_{2}{} ^{(2)}R^{\alpha\beta}\wedge{}
^{(4)}R_{\alpha\beta} \right. \nonumber\\ & & 
\qquad + \left.
{\mu}_{3}{}^{(3)}R^{\alpha\beta}\wedge{} ^{(6)}R_{\alpha\beta} +
{\mu}_{4}{} ^{(5)}R^{\alpha\beta}\wedge{} ^{(5)}R_{\alpha\beta}
\right).\label{PGgeneralL}
\end{eqnarray}
Here $R$ is the scalar curvature and $X$ is the pseudoscalar curvature ($X\equiv -\frac12R_{\alpha\beta\mu\nu}\eta^{\alpha\beta\mu\nu}$).  The torsion has been decomposed into three algebraically irreducible pieces: $T^\alpha={}^{(1)}T^\alpha+{}^{(2)}T^\alpha+{}^{(3)}T^\alpha$, which are, respectively, a pure tensor (16 components), the trace (vector), and a totally antisymmetric part (dual to an axial vector). Similarly the curvature 2-form has been decomposed into a sum of 6 algebraically irreducible pieces: $R^\alpha{}_\beta=\sum_{I=1}^6{}^{(I)}R^\alpha{}_\beta$, namely, in numerical order: weyl, pair-commutator, pseudoscalar, ricci-symmetric, ricci-antisymmetric, and scalar.  The respective number of components is (10,9,1,9,6,1).
In the above Lagrangian the parameters $\Lambda,a_0,a_I,w_I$ (which multiply even parity 4-forms) are scalars, and the parameters $b_0,\sigma_I,\mu_I$ (which multiply odd parity 4-forms) are pseudoscalars.
The general theory has 11 scalar plus 7 pseudoscalar parameters.  But they are not all physically independent.
They are subject to 1 even parity and 2 odd total differentials, leaving effectively 10 scalar + 5 pseudoscalar = 15 ``physical'' parameters, as we will briefly explain, referring to Refs.~\refcite{BHN,BH} for details.

\subsection{Topological terms}

Not all of the above parameters are physically independent, since there are 3 topological invariants.
Without changing the field equations, one can add to the Lagrangian 4-form~(\ref{PGgeneralL}) any multiple of the (odd parity) \emph{Nieh-Yan identity}~\cite{NY}:
\begin{equation}
 T^\alpha\wedge T_\alpha-R_{\alpha\beta}\wedge\vartheta^{\alpha\beta} \equiv  d(\vartheta^\alpha\wedge T_\alpha).
 \end{equation}
Also one can add a multiple of the (even parity) \emph{Euler 4-form}  $R^{\alpha\beta}\wedge R^{\gamma\delta}\eta_{\alpha\beta\gamma\delta}$.   Because of the 2nd Bianchi identity~(\ref{Bianchi2Id}), this makes no contributions to the field equations.
Furthermore once can add a multiple of the (odd parity) \emph{Pontryagin} 4-form $R^\alpha{}_\beta\wedge R^\beta{}_\alpha$. Again, thanks to the 2nd Bianchi identity, this has no effect on the field equations.
The actual physical equations will only depend on the 15 combinations of the 18 parameters that are invariant under such transformations.

The PG dynamics has been discussed in detail in Ref.~\refcite{GR100} including
(i) the Lagrangian, both 2nd and 1st order,
(ii) the Noether symmetries, conserved currents and differential identities,
(iii) the covariant Hamiltonian including the generators of the local Poincar\'e gauge symmetries,
(iv) our {preferred Hamiltonian boundary term},
(v) the quasi-local energy-momentum and angular momentum/center-of-mass moment obtained therefrom, and
(vi) the {choice of reference} in the boundary term.
We will include below a brief report of the Hamiltonian, its boundary term, and the associated quasi-local quantities.
The general PG homogeneous and isotropic cosmology has been considered by our group recently.\cite{Ho:2015ulu}

\section{The covariant Hamiltonian formulation}

\emph{Overview}: The Hamiltonian for dynamical geometry generates the evolution of a spatial region along a vector field. It includes a boundary term which determines both the value of the Hamiltonian and the boundary conditions. The value gives the quasi-local quantities: energy-momentum, angular-momentum and center-of-mass. The boundary term depends not only on the dynamical variables but also on their reference values; the latter determine the ground state (having vanishing quasi-local quantities). For our preferred boundary term for the PG (including Einstein's GR as a special case) we proposed 4D isometric matching and extremizing the energy to determine the reference metric and connection values.

Although the global \emph{total} energy-momentum is well defined (for spaces with suitable asymptotic regions), for any gravitating system --- and hence for all real \emph{physical} systems --- the localization of energy-momentum is still an outstanding fundamental problem.~\cite{Sza09} Unlike all matter and other interaction fields, the gravitational field itself has \emph{no} proper energy-momentum density. In view of the fact that energy-momentum is conserved, and that sources \emph{exchange} energy-momentum \emph{locally} with the gravitational field, one expects some kind of ``local description'' of the energy-momentum density of gravity itself. But all attempts at constructing such an expression led only to reference frame dependent quantities, generally referred to as \emph{pseudotensors}.~\cite{CNC99} Physically this can be understood as a consequence of \emph{Einstein's equivalence principle}: gravity cannot be detected at a point.
As mentioned earlier, Noether showed in her 1918 paper\cite{KS-Noether,Rowe99} that there is no proper energy density.
The energy-momentum of gravity --- and thus for all physical systems is inherently \emph{non-local}. The modern idea is \emph{quasi-local}: energy-momentum is associated with a closed surface bounding a region.~\cite{Penrose82}

From a first order Lagrangian formulation 
which gives pairs of first order equations for a $k$-form $\varphi$ and its conjugate $p$,  we developed a 4D-\emph{covariant\/} Hamiltonian formalism.\cite{CNT95,CN99,CNC99,CN00,CNT05,Nester08,GR100}
The Hamiltonian generates the evolution of a spatial region along a vector field.

The \emph{first order} Lagrangian for a \emph{$k$-form\footnote{This can include several fields of various types and grades with their indicies suppressed.}
 field $\varphi$} and its \emph{conjugate momentum $p$} is given by
\begin{equation}
{\cal L} = \mathcal{L}(d\varphi; \varphi, p) = d \varphi \wedge p - \Lambda(\varphi, p). \label{1st Lagrangian}
\end{equation}
The variation (with respect to $\varphi$ and $p$ independently)
\begin{equation}
\delta {\cal L} = d(\delta \varphi \wedge p) + \delta \varphi \wedge \frac{\delta {\cal L}}{\delta \varphi} + \frac{\delta {\cal L}}{\delta p} \wedge \delta p \label{deltaL}
\end{equation}
gives the equations of motion, with $\varsigma := (-1)^k$,
\begin{equation}
\frac{\delta {\cal L}}{\delta p} := d \varphi - \partial_p \Lambda = 0, \qquad \frac{\delta {\cal L}}{\delta \varphi} := - \varsigma d p - \partial_\varphi \Lambda = 0.
\end{equation}

Infinitesimal diffeomorphism invariance (in (\ref{deltaL}) replacing $\delta$ by the Lie derivative operator on forms: $\Lie_Z = d i_Z + i_Z d$), leads to an identity for \emph{any vector $Z$}:
\begin{equation}
d i_Z {\cal L} \equiv \Lie_Z {\cal L} \equiv d( \Lie_Z \varphi \wedge p) + \Lie_Z \varphi \wedge \frac{\delta {\cal L}}{\delta \varphi} + \frac{\delta {\cal L}}{\delta p} \wedge \Lie_Z p.
\end{equation}
From this one gets a ``translational current'' 3-form
which is conserved \emph{on shell\/} (i.e., when the field equations are satisfied):
\begin{eqnarray}
- d {\cal H}(Z) &\equiv& \Lie_Z \varphi \wedge \frac{\delta {\cal L}}{\delta \varphi} + \frac{\delta {\cal L}}{\delta p} \wedge \Lie_Z p,\label{dH}\\
{\cal H}(Z) &:=& \Lie_Z \varphi \wedge p - i_Z {\cal L}\\
&\equiv&\zeta i_Z\varphi\wedge dp+\zeta d\varphi\wedge i_Zp+i_Z\Lambda+d(i_Z\varphi\wedge p)\\
&=:& Z^\mu {\cal H}_\mu + d {\cal B}(Z).\label{ZH+dB}
\end{eqnarray}

A consequence of the expression (\ref{ZH+dB}) and (\ref{dH}) is
\begin{eqnarray}
&& d{\cal H}(Z) = d [ Z^\mu {\cal H}_\mu + d {\cal B}(Z) ] \equiv d Z^\mu \wedge {\cal H}_\mu + Z^\mu d {\cal H}_\mu,
\\
&&\Longrightarrow \qquad {\cal H}_\mu \; \textrm{vanishes ``on shell''}. \nonumber
\end{eqnarray}
Hence for gravitating systems the Noether translational ``charge'' --- \emph{energy-momentum} --- is \emph{quasi-local}, it is given by the integral of the boundary term, ${\cal B}(N)$:
\begin{equation}
E(Z,\Sigma)=\int_{\Sigma}{\cal H}(Z)=\oint_{\partial\Sigma}{\cal B}(Z).\label{E(Z)}
\end{equation}
But the total differential/boundary term can be completely modified:
\begin{equation}
\mathcal{H}' = \mathcal{H} + d \mathcal{B}' \quad \Longrightarrow \quad d \mathcal{H} = d \mathcal{H}'.
\end{equation}
This does not change the conservation property
 (it is an instance of the usual Noether current ambiguity---essentially adding a curl preserves the vanishing divergence property), but such a modification does change the conserved value.
The Hamiltonian approach \emph{tames this ambiguity}.

${\cal H}(Z)$ is not merely the Noether translational current, it is also the \textit{generator} of local diffeomorphisms, i.e., the \textit{Hamiltonian} density (3-form) which evolves a spacetime region along the vector field $Z$:
\begin{equation}
H(Z,\Sigma)=\int_\Sigma{\cal H}(Z)=\int_\Sigma Z^\mu{\cal H}_\mu  +\oint_{S=\partial\Sigma}{\cal B}(Z).\label{H(Z)}
\end{equation}
From this perspective one can identify the roles played by its separate pieces: the 3-form ${\cal H}_\mu Z^\mu$ generates the infinitesimal displacements along the vector field $Z$ (the Hamiltonian equations); this follows from the easily verified variational identity:
\begin{equation}
\delta {\cal H}(N) \equiv - \delta \varphi \wedge \Lie_N p + \Lie_N \varphi \wedge \delta p + d i_N(\delta \varphi \wedge p) - i_N \left( \delta \varphi \wedge \frac{\delta {\cal L}}{\delta \varphi} + \frac{\delta {\cal L}}{\delta p} \wedge \delta p \right).\label{deltaH}
\end{equation}

The boundary term, on the other hand, has two roles.
(i) The Hamiltonian value---the quasi-local quantities---as can be seen from (\ref{E(Z)}), (\ref{H(Z)}) are determined only by the surface integral:
\begin{equation}
E(Z,\Sigma) = \int_\Sigma {\cal H}(Z) = \int_\Sigma \left[ Z^\mu {\cal H}_\mu + d {\cal B}(Z) \right] = \oint_{\partial\Sigma} {\cal B}(Z).
\end{equation}
(ii)   Our Noether analysis revealed that ${\cal B}(N)$ can be adjusted, changing the conserved value to a new value. Fortunately the variational principle contains a (largely overlooked) feature which distinguishes all of these choices. The \emph{boundary variation principle}: the \emph{boundary term} in the \emph{variation} of the Hamiltonian (\ref{deltaH}) shows  what is to be held fixed on the boundary---thus it determines the boundary conditions~\cite{CNT95,CN99}. 
%
Hence the \emph{first ambiguity}---which expression?---is clearly under physical control: different Hamiltonian boundary term quasi-local expressions are associated with different types of physical boundary conditions.  This is similar to thermodynamics where there are various ``energies'' (internal, enthalpy, Helmholtz, Gibbs, etc.) which correspond to how the system interacts with the outside through its boundary.
 The Hamiltonian boundary term should be adjusted to give suitable boundary conditions.
%

 In general (in particular for gravity) it is necessary (in order to guarantee functional differentiability of the Hamiltonian on the phase space with the desired boundary conditions) to adjust the boundary term ${\cal B}(N) = i_N \varphi \wedge p$ which is naturally inherited from the Lagrangian~\eqref{1st Lagrangian}.

 We were led to a 2 parameter set of general boundary terms which are linear in $\Delta\varphi:=\varphi-\bar\varphi$, $\Delta p:= p-\bar p$, where $\bar\varphi,\bar p$ are reference values:~\cite{CNT95,CN99,CNT05,So:2006zz}
\begin{equation}
{\cal B}(Z) := i_Z\{a\varphi+(1-a)\bar\varphi\}
\wedge \Delta p - \zeta \Delta\varphi \wedge i_Z
\{bp+(1-b)\bar p\}.
 \label{genB}
\end{equation}
The associated variational Hamiltonian boundary term is
\begin{eqnarray}
\delta{\cal H}(Z) &\sim& d\Bigl[
\{ai_Z \delta\varphi \wedge \Delta p  - (1-a)i_Z \Delta\varphi \wedge \delta p\} \nonumber\\
&&\quad + \zeta\{
 -b\Delta\varphi \wedge i_Z \delta p +(1-b) \delta\varphi \wedge i_Z \Delta p \}
 \Bigr]. \label{deltaHZbound}
\end{eqnarray}
Here the extreme values $a,b=\{(0,0),(0,1),(1,0),(1,1)\}$ represent essentially a choice of Dirichlet (fixed field) or Neumann (fixed momentum) boundary conditions for the space and time parts of the fields separately.

For asymptotically flat spaces the Hamiltonian with any of these boundary term expressions is \emph{well defined}, i.e., the boundary term in its variation vanishes and the quasi-local quantities are well defined---at least on the phase space of fields satisfying Regge-Teitelboim\cite{Regge:1974zd}
 like asymptotic parity/fall-off conditions:
\begin{equation}
\Delta \varphi \approx {\cal O}^+(1/r) + {\cal O}^-(1/r^2), \qquad \Delta p \approx {\cal O}^-(1/r^2) + {\cal O}^+(1/r^3). \label{5:asymptotics}
\end{equation}
Also from (\ref{deltaH}), (\ref{deltaHZbound})  the formalism has natural energy flux expressions.\cite{CNT05}

The Hamiltonian boundary terms determines the values of the quasi-local quantities. For asymptotically flat spaces:
\begin{itemize}
  \item energy is given by a suitable \emph{timelike} displacement;

  \item linear momentum is obtained from a \emph{spatial} translation;

  \item angular momentum from a suitable \emph{rotational} displacement;

  \item a \emph{boost} will give the  center-of-mass moment.
\end{itemize}

\section{Application to the PG and GR}
A first order Lagrangian for Einstein's (vacuum) gravity theory is
\begin{equation}
{\cal L}_{\rm GR} = \frac1{2\kappa}R^\alpha{}_\beta \wedge \eta_\alpha{}^\beta,
\end{equation}
where $\eta^{\alpha\beta} := *(\vartheta^\alpha \wedge \vartheta^\beta)$. Our general formalism with $\varphi \to \Gamma^\alpha{}_\beta$ and $p \to \eta_\alpha{}^\beta$ gives a 2 parameter set of quasi-local expressions for GR:
\begin{equation}
2\kappa{\cal B}(Z) :=  \Delta \Gamma^\alpha{}_\beta \wedge i_Z
[a\eta_\alpha{}^\beta+(1-a){\bar\eta}{}_\alpha{}^\beta] + [bD_\beta Z^\alpha (1-b){\bar D}_\beta Z^\alpha] \Delta \eta_{\alpha}{}^\beta.
\end{equation}

\subsection{Preferred Hamiltonian boundary terms}

For GR,
 we identified a distinguished expression with some desirable properties:
\begin{equation}
{\cal B}_\vartheta(Z) := \frac1{2\kappa} \left( \Delta \Gamma^\alpha{}_\beta \wedge i_Z \eta_\alpha{}^\beta + {\bar D}_\beta Z^\alpha \Delta \eta_{\alpha}{}^\beta \right).\label{BprefGR(Z)}
\end{equation}
For this expression
\begin{equation}
\delta{\cal H}_{\vartheta}(Z)\simeq d i_Z(\Delta\Gamma^\alpha{}_\beta\wedge\delta\eta_\alpha{}^\beta).
\end{equation}
Hence it corresponds to imposing boundary conditions on a \emph{manifestly covariant object} (the coframe, i.e. essentially the metric---the obvious variable choice).
The associated energy flux expression is
\begin{equation}
\Lie_Z {\cal H}_{\vartheta} \simeq d i_Z \left( \Delta \Gamma^\alpha_\beta \wedge \Lie_Z \eta_\alpha{}^\beta \right).
\end{equation}

Like many other choices, for asymptotically flat spaces at spatial infinity, ${\cal B}_{\vartheta}(Z)$ (\ref{BprefGR(Z)}) gives the standard values for
energy-momentum and angular momentum/center-of-mass momentum.\cite{ADM,MTW,Regge:1974zd,Beig:1987zz,Szabados:2003yn}
Our preferred GR expression has some special virtues, including:
(i) at null infinity it gives the Bondi-Trautman energy and Bondi energy flux,\cite{CNT05}
%
(ii) it is covariant,
(iii) it can give positive energy,
%
%
%
%
(iv) for a suitable choice of reference it vanishes for Minkowski space.

For the PG our preferred Hamiltonian boundary term is
\begin{equation}
{\cal B}_{\rm PG}(Z) = i_Z \vartheta^\alpha \tau_\alpha + \Delta \Gamma^\alpha{}_\beta \wedge i_Z \rho_\alpha{}^\beta + {\bar {D}}_\beta Z^\alpha \Delta \rho_\alpha{}^\beta. \label{BprefPG(Z)}
\end{equation}
For more details about it please see the works cited above.

\section{The Reference Choice}

Regarding the \emph{second ambiguity} inherent in our quasi-local energy-momentum expressions: \emph{the choice of reference}.  In principle, one could use any reference appropriate to the physical application. In practice one normally wants a very symmetrical reference.
 Only if the chosen reference is a space of \emph{constant curvature} (positive for de Sitter, negative for anti-de Sitter and zero for Minkowski) will one have 10 reference Killing vector fields that can be used for the vector $Z$ to define 10 quasi-local quantities via (\ref{BprefGR(Z)}) and (\ref{BprefPG(Z)}) for GR and the PG, respectively. For a Minkowski reference they are the energy-momentum, angular momentum and center-of-mass moment.

Here we present some details just for the case of GR with a Minkowski reference.
 One then needs to choose a specific Minkowski space.  Recently we proposed
(i) \emph{4D isometric matching on the boundary},\footnote{The hardest part of 4D isometric matching is the embedding of the 2D surface $S$ into Minkowski space; Yau and coworkers have extensively investigated this.\cite{WangYau}}
and (ii) \emph{energy optimization\/}
as criteria for the ``best matched'' reference on the boundary of the quasi-local region.
 This proposal has been tested on spherically symmetric and axisymmetric spacetimes.\cite{SCLN15}

    Essentially one needs a reference geometry in the neighborhood of the boundary of the region.  One could view this as embedding a neighborhood of  the 2-boundary into Minkowski space.\cite{WCLN}
One construction is to choose, in a neighborhood of the desired spacelike boundary 2-surface $S$, 4 smooth functions $y^i = y^i(x^\mu), \; i = 0, 1, 2, 3$ with $dy^0 \wedge dy^1 \wedge dy^2 \wedge dy^3 \ne 0$ and use them to define a Minkowski reference:
\begin{equation}
\bar g = -(dy^0)^2 + (dy^1)^2 + (dy^2)^2 + (dy^3)^2.
\end{equation}
The reference connection is then
\begin{equation}
\bar \Gamma^\alpha{}_\beta = x^\alpha{}_i ( \bar\Gamma^i{}_j y^j{}_\beta + dy^i{}_\beta ) = x^\alpha{}_i dy^i{}_\beta,
\end{equation}
where $dy^i = y^i{}_\alpha dx^\alpha$ and $dx^\alpha = x^\alpha{}_j dy^j$ have been used along with vanishing Minkowski reference connection coefficients, $\bar \Gamma^i{}_j=0$. If $Z^\mu$ is a \emph{translational Killing field} of the Minkowski reference, then $\bar DZ$ vanishes, and hence so does the 2nd quasi-local term. Our quasi-local expression then takes the simpler form:
\begin{equation}
{\mathcal B}(Z) = Z^k x^\mu{}_k (\Gamma^\alpha{}_\beta - x^\alpha{}_j \, dy^j{}_\beta) \wedge \eta_{\mu\alpha}{}^\beta.
\end{equation}

How we determine the reference choice $y^{i}{}_{\mu}$ can be simply explained with the aid of quasi-spherical foliation adapted coordinates $t, r, \theta, \phi$.  Isometric matching on the 2-surface implies
\begin{equation}
g_{AB} = \bar g_{AB} = \bar g_{ij} y^i_A y^j_B = - y^0_A y^0_B + \delta_{ab} y^a_A y^b_B; \ \  a, b = 1, 2, 3; \; A, B = 2, 3 = \theta, \phi,
\end{equation}
where the reference metric on the dynamical space has the components $\bar g_{\mu\nu} = \bar g_{ij} y^i{}_\mu y^j{}_\nu$. From a classic closed 2-surface into $\mathbb R^3$ embedding theorem---as long as one restricts $S$ and $y^0(x^A)$ such that on $S$
\begin{equation}
g_{AB}' := g_{AB} + y^0_A y^0_B
\end{equation}
is convex---one has a unique isometric embedding. (But, unfortunately, there is no explicit formula.)

\subsection{4D Isometric Matching}
Complete 4D isometric matching on $S$ has 10 constraints:~\cite{ae100,cjp13,SCLN15,Sun:2016hbw}
\begin{equation}
g_{\mu\nu}|_S = \bar g_{\mu\nu}|_S = \bar g_{ij} y^i{}_\mu y^j{}_\nu|_S.
\end{equation}
(With this condition $\Delta\eta_\alpha{}^\beta$ vanishes, so the 2nd term in our quasi-local expression (\ref{BprefGR(Z)}) vanishes for all $Z$.)

There are 12 embedding functions on the constant $t, r$ 2-surface:
\begin{equation}
y^i (\Longrightarrow y^i_\theta, y^i_\phi), \quad y^i_t, \quad y^i_r.
\end{equation}
The 10 constraints split into 3 for the already discussed 2D isometric matching: $g_{\theta\theta}, g_{\theta\varphi}, g_{\varphi\varphi}$ which constrain the 4 $y^i$; 3 normal bundle algebraic quadratic expressions: $g_{tt}, g_{tr}, g_{rr}$; and 4 mixed linear algebraic expressions: $g_{t\theta}, g_{t\varphi}, g_{r\theta}, g_{r\varphi}$. The 2D isometric matching can be regarded as a given $y^0$ uniquely determining $y^1,y^2,y^3$ on $S$. The remaining 7 algebraic equations can be regarded as fixing the other 7 embedding variables in terms of $y^i$ and $y^0{}_r$. Thus one can take $y^0,\ y^0{}_r$ as the embedding control variables. Geometrically $y^0{}_r$ controls a boost in the normal plane.

An alternative approach is to regard 4D isometric matching in terms of orthonormal frames.  The reference geometry will have a reference frame of the form $\bar\vartheta^i=dy^i$.  If there is 4D isometric matching then one can choose the dynamical frame such that it can be Lorentz transformed to match such a reference frame at all points on the boundary:
\begin{equation}
L^i{}_\alpha(p)\vartheta^\alpha(p)=\bar\vartheta^i(p), \qquad \forall p\in S.
\end{equation}
Then these  2-forms restricted to $S$ must satisfy the integrability conditions:
\begin{equation}
d(L^i{}_\alpha\vartheta^\alpha)|_S=0.
\end{equation}
This is 4 2-forms on a 2D space, each 2-form having one component, thus this is 4 restrictions on the 6 parameters of the Lorentz transformation, so again we see that 4D isometric matching has 2 degrees of freedom.

\subsection{An Optimal Choice}
How to fix the remaining 2 degrees of freedom in the reference choice?
One can regard the value of the boundary term as a measure of the difference between the dynamical boundary values and the reference boundary values.\cite{ae100,cjp13,SCLN15,Sun:2016hbw}

For a given $S$ there are 2 related quantities which can be considered: $m^2 = - \bar g{}^{ij} p_i p_j=p_0^2-p_1^2-p_2^2-p_3^2$ and $E(Z, S)$.
The \emph{critical points} are distinguished.
Consider first finding the critical points of $m^2$.
Technically this is rather complicated: one would be extremizing a linear combination of quadratic quantities, each being an integral over $S$.  It would determine the reference only up to Poincar\'e transformations.  However, once could then use the available Lorentz ``gauge'' freedom to specialize to the reference in which the linear momentum vanished: $\vec p=0$.  In this ``center-of-momentum'' frame $m^2$ reduces to $p_0^2$.  But the critical points of $p_0^2$ are also  critical points of $p_0$.
  Thus one can arrive at the same reference by considering the much more simple case of the critical points of  $E(\partial_{y^0}, S)$.

Based on some physical and practical computational arguments, it is reasonable to expect that one could find a unique solution.
Consider being given a set of data from a numerical relativity calculation. Compute the energy given by a large number of reference choices, the critical values will stand out.
For our quasi-local values for axisymmetric solutions including Kerr we were able to explicitly find the critical point analytically.\cite{cjp13}

\section{Summary}
For any gravitating system --- and hence for all \emph{physical} systems --- the \emph{localization} of energy-momentum is an outstanding problem. We have discussed the relation between the covariant Hamiltonian boundary term, the quasi-local quantities and the physical boundary conditions.
For gravitating systems, using variables appropriate to a gauge theory perspective, we found certain quasi-local energy-momentum expressions; each is associated with a physically distinct, geometrically clear, boundary condition. We identified certain preferred expressions for the PG and GR. With a 4D isometric ``\emph{best matched}'' reference, we have a method to determine the Hamiltonian boundary term quasi-locally for locally Poincar\'e gauge invariant gravity including GR. This in particular gives a way of resolving the ambiguities in determining the quasi-local energy-momentum and angular momentum of classical physical systems.


\section*{Acknowledgments}

C.M.C. was supported by the Ministry of Science and Technology of the R.O.C.
under the grant MOST 106-2112-M-008-010.

\end{document}